\documentclass[preprint,preprintnumbers,amsmath,amssymb,nofootinbib]{revtex4}

\usepackage{graphicx}
\usepackage{dcolumn}
\usepackage{bm}

\begin{document}


\title{Universal ratio as lower bound}

\author{Wen-Yu Wen}
\affiliation{Department of Physics and Center for Theoretical
Sciences\\ National Taiwan University, Taipei 106, Taiwan}
 \email{steve.wen@gmail.com}


\begin{abstract}
We refine the definition of universal ratio $c/\tilde{c}$ obtained via AdS/CFT correspondence as shown in \cite{Kovtun:2008kw} , denoted as $\gamma_c$, and apply it to non-CFTs whose dual gravitational theory have metric of asymptotic AdS.  We conjecture that $\gamma_c=1$ is the lower bound being saturated at high temperature regime and serves as an ordering parameter as energy scale varies.  We test this conjecture with the hard wall AdS/QCD toy model and $N=2^*$ Pilch-Warner solution and find the agreement.  At last, we make a connection with the $\cal C$-Theorem and prove this conjecture in a general holographic background with multi-kink geometry.
\end{abstract}


\maketitle

\section{Introduction}
In the paper \cite{Kovtun:2008kw} the authors argued that a universal ratio $c/\tilde{c}$ exists in an infinite class of conformal field theories in diverse dimensions, which possess a classical gravity dual according to the AdS/CFT correspondence\cite{Maldacena:1997re,Gubser:1998bc,Witten:1998qj}.  Here $c$ is the central charge and $\tilde{c}$ is the normalized entropy density.  It may serves as a criterion for whether a given CFT has a dual gravitational description.  In this note, we refine the definition of universal ratio, denoted as $\gamma_c$, so that it can be applied to those no-CFTs whose dual gravitational theory has metric of asymptotic AdS.  We argue that $\gamma_c=1$ is the lower bound which is saturated at high temperature (UV) regime.  It also serves as an ordering parameter in different energy scales.   Later on, we test this conjecture with the hard wall AdS/QCD toy model and $N=2^*$ Pilch-Warner (PW) solution.  In the former case, $\gamma_c$ is found to jump between $1$ and $\infty$ at some critical temperature signaling a first-order phase transition.  In the PW solution, we expect $\gamma_c$ increases smoothly from its lower bound as temperature decreases.  Both saturate $\gamma_c=1$ at their UV limit.  At last, we make a connection with the $\cal C$-Theorem and prove the conjecture by showing that the monotonically increasing $\gamma_c$ in a general holographic background with multi-kink geometry.

\section{refined universal ratio $\gamma_c$}
We first recall the definition of universal ratio $c/\tilde{c}$\cite{Kovtun:2008kw}.  The central charge $c$ is defined via two-point function, which can be computed via AdS$_{d+1}$/CFT$_d$ prescription\cite{Gubser:1998bc,Witten:1998qj,Freedman:1998tz} and has the following form
\begin{equation}\label{eq:c_charge}
c=\frac{d+1}{d-1}\frac{L^{d-1}}{4\pi G_N^{(d+1)}}\frac{\Gamma(d+1)\pi^{d/2}}{\Gamma(d/2)^3}.
\end{equation}
On the other hand, the renormalized entropy density $\tilde{c}$ is defined via 
\begin{equation}
s=\tilde{c}\frac{\Gamma(d/2)^3}{4\pi^{d/2}\Gamma(d)}(\frac{4\pi}{d})^d\frac{d-1}{d+1}T^{d-1},
\end{equation}
where the entropy density s can be computed for AdS Schwarzschild black hole, that is
\begin{equation}
s=\frac{1}{4G_N^{(d+1)}}(\frac{4\pi L}{d})^{d-1}T^{d-1}.
\end{equation}
In order to carry this definition over to non-CFTs, we request that the dual gravitational description has metric of asymptotic AdS such that the corresponding field theory approaches CFT at UV limit.  Then we can refine $c$ as the short distance limit of two-point function, which is still given by (\ref{eq:c_charge}).  However, the nontrivial variation of degrees of freedom at different energy scales give rise to $\tilde{c}\neq c$ and therefore the ratio $\gamma_c \neq 1$.  Now it is obvious that $\gamma_c=1$ is saturated at UV limit by assumption, however it is nontrivial statement that $\gamma_c\ge 1$ as we are away from UV regime.  We first give a hand-waving argument by observing another related ratio of shear viscosity to entropy density $\eta/s$.  This quantity can be obtained from the ratio of absorption cross-section to entropy density, $\sigma/s$ , in the dual gravitational description, and conjectured to be universal, say $\eta/s = 1/4\pi$ for both CFTs and their relevant deformations\cite{Buchel:2003tz}.  Suppose that we have some kind of thermal object $X$\footnote{More likely X is a compound object, such as a black hole couples with other matter fields.} inside the AdS$_{d+1}$ space in correspondence to certain non-CFT at finite temperature on its boundary, in similarity as the AdS-Schwarzschild black hole to CFT.  Since black holes have the largest cross-section as well as entropy density in comparison to other objects of the same size, we expect that $\sigma_X < \sigma_{BH}$.  If we agree with the universality of the ratio $\sigma/s$, then we must also have $s_X< s_{BH}$, hence the normalized entropy density $\tilde{c}_X < \tilde{c}_{BH}$.  In this way, we deduce that $\gamma_c>1$ for non-CFTs.  In the following we will examine two examples of non-CFTs and find they all agree with our conjecture that $\gamma_c \ge 1$.

\section{hard wall AdS/QCD toy model}
The hard wall toy model was constructed via bottom-up approach and has been proved useful for calculating hadronic spectrum in QCD\cite{BoschiFilho:2002ta,deTeramond:2005su,Erlich:2005qh,Da Rold:2005zs}.  It consists of the Poincar\'{e} AdS space with a hard infrared cut off at $z=z_0$, that is\cite{Polchinski:2001tt}
\begin{equation}
ds^2=\frac{r^2}{L^2}(-dt^2+dx_{d-1}^2)+\frac{L^2}{r^2}dr^2,\qquad 0\le z<z_0.
\end{equation} 
It was shown that in this model Hawking-Page thermal phase transition can still happen at a finite critical temperature regardless of its noncompact boundary, where the critical temperature is obtained by $T_c=2^{1/4}/\pi z_0$\cite{Herzog:2006ra}.  The cut-off thermal AdS is thermodynamically favorable at the temperature lower than $T_c$, while the black hole is favorable at the temperature higher than $T_c$.  With appropriate choice of counterterm which is required for finite gravity action, the computation of entropy in both cases has been carried out in the , that is\cite{BallonBayona:2007vp}

\[s = \left\{ 
\begin{array}{l l}
  0 & \quad \mbox{if $T<T_c$}\\
  \frac{\pi^2N^2}{2}T^3 & \quad \mbox{if $T>T_c$}\\
\end{array} \right. \]

Provided the five dimensional Newton constant $G_N^{(5)}L^5=8\pi^3 g^2\alpha'^4$ and the relation $L^4=4\pi gN\alpha'^2$, we identify that there exists a first order phase transition at $T_c$ where $\gamma_c$ jumps as we go from UV limit to IR limit, that is

\[\gamma_c = \left\{ 
\begin{array}{l l}
  \infty & \quad \mbox{if $T<T_c$}\\
  1 & \quad \mbox{if $T>T_c$}\\
\end{array} \right. \]

This sudden change of degrees of freedom is also expected in the soft wall toy model, where the IR wall is provided by an exploding dilaton at IR regime\cite{BallonBayona:2007vp}. 

\section{$N=2^*$ PW solution}
In the construction of PW solution\cite{Pilch:2000ue}, ${\cal N}=4$ super Yang-Mills is softly broken into ${\cal N}=2$ by introducing a mass for the adjoint hypermultiplet.  The supergravity background dual to its finite temperature version was studied in \cite{Buchel:2003ah,Buchel:2004hw,Buchel:2007vy}.  It consists of two scalars $\alpha$ and $\chi$, which in the field theory side correspond to operators associated with boson and fermion mass individually, satisfying the five dimensional equations of motion,
\begin{eqnarray}
&&\frac{1}{4}R_{\mu\nu}=3\partial_\mu \alpha \partial_\nu \alpha + \partial_\mu \chi \partial_\nu \chi + \frac{1}{3}g_{\mu\nu}{\cal P},\nonumber\\
&&\Box \alpha =\frac{1}{6}\frac{\partial {\cal P}}{\partial \alpha},\qquad \Box \chi = \frac{1}{2}\frac{\partial {\cal P}}{\partial \chi},
\end{eqnarray}
where ${\cal P}={\cal P}(\alpha,\chi)$ is the effective scalar potential whose explicit form is not important here.  The thermodynamical quantities apart from the high temperature regime, say $\frac{m}{T}\ll 1$\footnote{Here $m$ is the deformation parameter which gives the mass term $\frac{m}{g_{YM}^2}(Tr Q^2 + Tr \tilde{Q}^2)$ for chiral multiplets $Q$ and $\tilde{Q}$.}, were computed and in particular, the leading high-temperature correction of entropy function was given by\cite{Buchel:2004hw},
\begin{equation}
s\simeq \frac{\pi^2 N^2}{2}T^3 (1-\frac{\Gamma(3/4)^4}{\pi^4}\frac{m^2}{T^2}).
\end{equation}  
This immediately implies that our refined universal ratio takes the following form at high temperature regime,
\begin{equation}
\gamma_c \simeq 1+ \frac{\Gamma(3/4)^4}{\pi^4}\frac{m^2}{T^2}.
\end{equation}
This can be generalized to the non-supersymmetric case where bosonic and fermionic mass are not equal, say $m_b \neq m_f$\cite{Buchel:2007vy}, then one can derive that
\begin{equation}
\gamma_c \simeq 1+ \frac{48}{\pi^2}\frac{1}{(24\pi)^2}\frac{m_b^4}{T^4}  +\frac{4}{\pi}\frac{\Gamma(3/4)^4}{\pi^4}\frac{m^2}{T^2}.
\end{equation}
In either case we observe that $\gamma_c\ge 1$ as we move away from UV.

\begin{figure}\label{fig_1}
\includegraphics[width=0.49\textwidth]{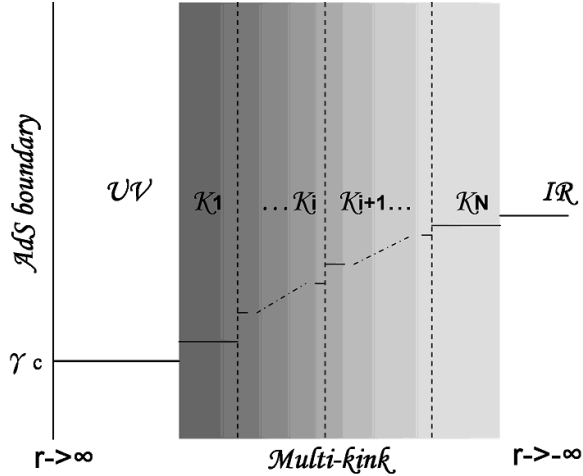}
\caption{Asymptotic AdS space with multi-kink geometry.  The refined universal ratio $\gamma_c$ is monotonically increasing from UV to IR.}
\end{figure}

\section{Discussion}
What we have shown is that the $\gamma_c \ge 1$ works for both hard wall AdS/QCD toy model and $N=2^*$ PW solution, where $\gamma_c=1$ serves as a lower bound at UV regime.  Here we would like to make a connection to the $\cal C$-Theorem\cite{Freedman:1999gp} such that a monotonically decreasing $\cal C$ implies a monotonically increasing $\gamma_c$.  Consider a general possible $d+1$-dimensional bulk metric where the Poincar\'{e} symmetry is respected on the boundary, that is,
\begin{equation}
ds^2=e^{2A(r)}\eta_{\mu\nu}dx^{\mu}dx^{\nu}+dr^2
\end{equation}
We will take $A(r)\to r/L$ for asymptotic AdS space of curvature radius $L$ as $r\to\infty$.  The infrared region is given by $r\to -\infty$.  The $\cal C$-Theorem states that there exists a function\cite{Freedman:1999gp}
\begin{equation}
{\cal C}(r)=\frac{{\cal C}_0}{A'(r)^{d-1}},
\end{equation}
which is non-increasing along the flow toward the IR provided the weaker energy condition saying that $A''(r)\le 0$.  Since we only concern positive $\cal C$ in arbitrary dimensions, we also request $A'(r)\ge 0$.  One typical example is the kink ansatz given in \cite{Freedman:1999gp} where AdS space appears in both UV and IR regions and a kink interpolates in between.  We now propose a multi-kink ansatz where each kink of size $\Delta$ is labeled by increasing integer from UV to IR in the following order $UV|{\cal K}_1|{\cal K}_2|\cdots|{\cal K}_N | IR$ as shown in the Fig.$1$.  $A(r)$ behaves linearly inside each kink ${\cal K}_i$, say $A(r)=(r-a_i)/L_i+b_i$ for $r\in {\cal K}_i$ and constant $a_i,b_i$.  The discontinuity of $A'(r)$ at each kink interface may be further smoothed out with care by some transition function.  Once we heat up the system, we expect the blacken function locally behaves like $f(r)=1-e^{-\frac{d}{L_i}(r-r_0)}$ as the horizon $r_0\in {\cal K}_i$.  The entropy density can be computed as follows,
\begin{equation}
s=\frac{1}{4G_N^{(d+1)}}(\frac{4\pi L}{d})^{d-1}T^{d-1}(\frac{d}{Lf'(r_0)})^{d-1}.
\end{equation}
This implies that the refined universal ratio in each kink takes the following form
\begin{equation}
\gamma_c^i = (\frac{L_i}{L})^{1-d}.
\end{equation}
The monotonically increasing $A'(r)$ implies that $L_{UV}>\cdots>L_i>L_{i+1}>\cdots>L_{IR}$, therefore 
\begin{equation}
1=\gamma_c^{UV}<\cdots<\gamma_c^{i}<\gamma_c^{i+1}<\cdots<\gamma_c^{IR}
\end{equation}
Once we take the continuous limit where kink size $\Delta\to 0$ and $N\to \infty$, we may obtain a smooth and non-decreasing $\gamma_c$ along the flow toward the IR, which acts in opposite direction to the function $\cal C$.  This concludes our proof for $\gamma_c\ge 1$ in a general holographic background with asymptotic AdS.

\begin{acknowledgments}
The author is grateful to NTU Physics department for presenting this work in the Colloquium at its early stage.  I would like to thank Pei-Ming Ho and Eiji Nakano for useful comments.  This work is partially supported by the Taiwan's National Science Council under Grant No. NSC96-2811-M-002-018 and NSC97-2119-M-002-001.
\end{acknowledgments}

\bibliography{apssamp}

\end{document}